\documentstyle[amssymb,aps]{revtex}

\begin{document}

\title{Exact Density Functionals in One Dimension}

\author{Joachim Buschle, Philipp Maass, and Wolfgang Dieterich}

\address{Fakult\"at f\"ur Physik, Universit\"at Konstanz, D-78457 Konstanz,
  Germany}

\date{16 November 1999}

\maketitle

\begin{abstract}
We propose a new and general method for deriving exact density
functionals in one dimension for lattice gases with finite-range
pairwise interactions. Corresponding continuum functionals are
derived by applying a proper limiting procedure. The method is based
on a generalised Markov property, which allows us to set up a rather
transparent scheme that covers all previously known exact
functionals for one-dimensional lattice gas or fluid systems.
Implications for a systematic construction of approximate density
functionals in higher dimensions are pointed out.
\end{abstract}

\vspace*{1.5cm}
Density functional theory provides one of the most powerful methods
for studying thermodynamic properties of inhomogeneous fluids
\cite{Oxtoby:1991,Evans:1992,Percus:1994,Loewen:1994}.  Various
schemes are known to approximate the excess part of free energy
functionals, which contains the contribution from the interactions
between the molecules. Exact density functionals, however, have only
been derived in one dimension $d=1$. These functionals have their
merits in establishing ``boundary conditions'' in the sense that
approximate functionals used in $d>1$ should become exact upon
dimensional reduction to $d=1$. In this Letter we will present a
simple general method how to derive exact density functional in one
dimension for systems of particles interacting via a finite-range pair
potential.  In particular we will show that density functionals
calculated previously
\cite{Robledo/Varea:1981,Buschle/etal:2000,Percus:1976,Vanderlick/etal:1989,Percus:1982a,Percus:1989,Percus:1990,Percus/Zhang:1988,Zhang/Percus:1989,Samaj/Percus:1993}
can be recovered. We will first consider lattice systems from which
the corresponding continuum functionals can be straightforwardly
derived by a proper limiting procedure.

The lattice Hamiltonian reads
\begin{equation}
{\cal H}=\sum_s x_s\tilde u_s + \sum_{i<j} v_{i,j}\,x_i\,x_j\,,
\label{h-eq}
\end{equation}
where $v_{i,j}$ is the interaction potential, $\tilde u_s\!=\!u_s-\mu$
is the external potential minus the chemical potential $\mu$, and
$x_s$ are the occupation numbers ($x_s\!=\!1$ if site $s$ is occupied
and $x_s=0$ else). A particularly simple situation arises for a system
composed of hard rods of length $2m$ that is described by an
interaction potential $v_{i,j}\!=\!\infty$ for $j-i\!<\!2m$ and
$v_{i,j}\!=\!0$ else.  For this model an exact density functional was
derived by Robledo and Varea \cite{Robledo/Varea:1981} (see also
\cite{Buschle/etal:2000}). The corresponding continuum hard rod fluid
has been treated by Percus, who succeeded to set up the first exact
functional for this many-particle system with athermal interactions
\cite{Percus:1976}. Later exact functionals have been derived for the
multicomponent hard-rod \cite{Vanderlick/etal:1989} and sticky core
fluid \cite{Percus:1982a}, as well as for the Takahashi model
\cite{Percus:1989}.  The perhaps most general functional was set up
for a lattice system as in (\ref{h-eq}) with an arbitrary interaction
potential exhibiting a finite range \cite{Percus:1990}
($v_{i,j}\!=\!0$ for $j-i\!\ge\! l$).  This system, that we will refer
to as the ``finite-range model'' in the following, includes as a
special case the Ising model for which various exact functionals have
been derived separately for periodic boundary conditions
\cite{Percus/Zhang:1988,Zhang/Percus:1989,Samaj/Percus:1993}.

We will focus here on the finite-range model with fixed boundary
conditions, that means we consider a finite lattice of $M$ sites
($s\!=\!1,\ldots,M$) and fix the occupation numbers at the boundary
points $s\!=\!0$ and $s\!=\!M+1$ to be $x_0\!=\!x_{M+1}\!=\!1$.  Our
basic idea for deriving the density functional is to generate the
joint probability distribution $\chi[x_1,\ldots,x_M]$ by a generalised
Markov process, and then to equate the resulting expression with the
Boltzmann formula $\chi[x_1,\ldots,x_M]={\cal Z}^{-1}\exp(-{\cal H})$
(for convenient notation we set $k_{\rm B}T=1$).

Let us start by expressing the joint probability distribution in terms of
the conditional probabilities $\omega_s[x_s|x_{s-1},\ldots,x_1]$,
\begin{equation}
\label{chi-eq}
\chi[x_1,\ldots,x_M]=\prod_{s=1}^{M}\omega_s[x_s|x_{s-1},\ldots,x_1]\,.
\end{equation}
Due to the finite range of the interaction, the conditional
probabilities fulfil the Markov property
$\omega_s[x_s|x_{s-1},\ldots,x_1]=\omega_s[x_s|x_{s-1},\ldots,x_{s-(l-1)}]$
(\cite{markov-comm}; we set $x_s\!=\!0$ for $s\!<\!0$ for convenient
notation). Factorising $\omega_s[x_s|x_{s-1},\ldots,x_{s-(l-1)}]$ into
$2^l$ terms corresponding to the possible configurations
of $(x_s,\ldots,x_{s-(l-1)})$ yields
\begin{eqnarray}
\label{omega-eq}
\omega_s[x_s|x_{s-1},\ldots,x_{s-(l-1)}]&=&
\prod_{r=0}^{l-1}\prod_{{\,\mathbf j}\in{\cal A}_{s,r}}
\left[\left(\frac{a_s^{\,\mathbf j}}{b_s^{\,\mathbf j}}\right)^{x_s}
\left(1\!-\!\frac{a_s^{\,\mathbf j}}{b_s^{\,\mathbf j}}\right)^{1-x_s}
\right]^{\textstyle\prod\limits_{k=1}^r x_{j_k}
\prod\limits_{i\in I_{s,r}^{\,\mathbf j}}(1\!-\!x_i)}\,,\\
&&\hspace{-2.7cm}a_s^{\,\mathbf j}\equiv {\mathbb E}[x_s\prod_{k=1}^r x_{j_k}
   \prod_{i\in I_{s,r}^{\,\mathbf j}}(1-x_i)]\,,\hspace*{0.8cm}
b_s^{\,\mathbf j}\equiv {\mathbb E}[\prod_{k=1}^r x_{j_k}
\prod_{i\in I_{s,r}^{\,\mathbf j}}(1-x_i)]\,.\nonumber
\end{eqnarray}
${\mathbb E}[\ldots]$ denotes an expectation value with respect to
$\chi$ and ${\cal A}_{s,r}$ is the set of all configurations
$(x_{s-(l-1)},\ldots,x_{s-1})$ with exactly $r$ sites being occupied.
The superscript ${\mathbf j}$ specifies that particular element of
${\cal A}_{s,r}$, for which the $r$ sites $j_1,\ldots,j_r$ are
occupied, and $I_{s,r}^{\,\mathbf j}$ is the corresponding set of all
the remaining vacant sites.  Note that $a_s^{\,\mathbf
  j}/b_s^{\,\mathbf j}$ is the probability that the site $s$ is
occupied under the condition that the $r$ sites $j_1,\ldots,j_r$ are
occupied while the remaining sites $i\in I_{s,r}^{\,\mathbf j}$ are
empty. Although eq.~(\ref{omega-eq}) looks complicated, its structure
becomes immediately obvious when considering the special case $l=2$,
where $\omega_s[x_s|x_{s-1}]=\omega_s[1|1]^{x_sx_{s-1}}
\omega_s[1|0]^{x_s(1\!-\!x_{s-1})}\omega_s[0|1]^{(1\!-\!x_s)x_{s-1}}
\omega_s[0|0]^{(1\!-\!x_s)(1\!-\!x_{s-1})}$ with
$\omega_s[1|1]={\mathbb E}[x_sx_{s-1}]/{\mathbb E}[x_{s-1}]$ etc.

A grand canonical functional $\Omega\equiv-\log{\cal Z}$ is now
derived by inserting eq.~(\ref{omega-eq}) into eq.~(\ref{chi-eq}),
and by using for $\chi$ the Boltzmann formula.  After taking the
logarithm and the expectation value on both sides of
eq.~(\ref{chi-eq}), we obtain
\begin{equation}
\label{func1-eq}
\Omega=\sum_{s=1}^M p_s\tilde u_s+
\hspace*{-0.4cm}\sum_{0\le i<j\le M+1}
\hspace*{-0.5cm}v_{i,j}\,p_{i,j}+
\sum_{s=1}^M\sum_{r=0}^{l-1}\sum_{{\,\mathbf j}\in{\cal A}_{s,r}}\left[
a_s^{\,\mathbf j}\log\left(\frac{a_s^{\,\mathbf j}}{b_s^{\,\mathbf j}}\right)
+(b_s^{\,\mathbf j}\!-\!a_s^{\,\mathbf j})
\log\left(1\!-\!\frac{a_s^{\,\mathbf j}}{b_s^{\,\mathbf j}}\right)\right]
\nonumber
\end{equation}
where for $r\ge1$ and $k_1\!<\!\ldots\!<\!k_r$ we define
$p_{k_1,\ldots,k_r}\equiv {\mathbb E} [x_{k_1}\ldots x_{k_r}]$.
According to the definitions of $a_s^{\,\mathbf j}$ and
$b_s^{\,\mathbf j}$ (see eq.~(\ref{omega-eq})), $\Omega$ is a
functional of all moments $p_{k_1,\ldots,k_r}$ (with $k_r-k_1<l$),
that means it can be considered as a functional of (local)
correlations up to order $l-1$.  In order to derive a density
functional that depends only on the occupation probabilities $p_s$,
the correlations have to be expressed in terms of the $p_s$. According
to the theorem of Mermin \cite{Mermin:1965}, this can be done in a
manner that the corresponding ``correlator equations'' are invariant
with respect to $\tilde u_s$ and we will show next that there exists a
general method to achieve this \cite{naive-comm}.

To illustrate the method, we choose the special interaction potential
\begin{equation}
\label{v-eq}
v_{i,j}=\left\{
\begin{array}{c@{\hspace{0.6cm}}l}
\infty & 0\,,\le j-i<2m, \\[0.2cm]
\hat v_{i,j}\,,& 2m\le j-i\le l-1<4m\,.
\end{array}\right.
\end{equation}
This potential describes hard rods with lengths $2m$, which interact
via the potential $\hat v_{i,j}$, and it corresponds to a lattice
version of the continuum Takahashi model (for $\hat v_{i,j}=v_{j-i}$,
\cite{Takahashi:1942}).  The special form of $v_{i,j}$ implies that in
an interval of size $l\!-\!1$ there can at most two occupation numbers
be equal to one. Accordingly, $a_k^{\,\mathbf j}$ and $b_k^{\,\mathbf
  j}$ depend only on the occupation probabilities $\{p_s\}$ and the
two-point correlations $\{p_{r,s}\}$.
 
In order to calculate the two-point correlations, we consider the
following three cases: {\it (i)} all $x_s$ ($s\!=\!1,\ldots,M$) are
zero, {\it (ii)} exactly one of the $x_s$ is equal to one (and the
rest equal to zero), and {\it (iii)} exactly two occupation numbers
$x_i$, $x_j$ with $2m\!\le\!j-i\!<\!l$ are equal to one. By expressing
the probabilities for these cases in terms of the Boltzmann formula,
we find {\it (i)} $\chi[{\rm all}\;x_s\!=\!0]\equiv\chi[{\bf 0}]={\cal
  Z}^{-1}$, {\it (ii)} $\chi[x_s\!=\!1;{\rm
  rest}\;0]\equiv\chi[x_s\!=\!1;{\bf 0}]={\cal Z}^{-1} \exp(-\tilde
u_s)$, and {\it (iii)} $\chi[x_i\!=\!1,x_j\!=\!1;{\bf 0}]={\cal
  Z}^{-1}\exp[-\tilde u_i-\tilde u_j-\hat v_{i,j}]$ \cite{bound-comm}.
On the other hand, by considering eqs.~(\ref{chi-eq},\ref{omega-eq})
for the cases {\it (i)-(iii)} we obtain expressions containing the
sets $\{p_s\}$ and $\{p_{i,j}\}$.  For example, $\chi[{\bf
  0}]=\prod_{s=1}^M(1\!-\!a_s^{(0)}/b_s^{(0)})$.  The correlator
equations then follow by multiplying the expressions for the cases
{\it (i)-(iii)} in such a way that ${\cal Z}$ and $\{\tilde u_s\}$
cancel, i.e. in the present situation we get $\chi[{\bf
  0}]\chi[x_i=1,x_j=1;{\bf 0}]= \exp(- \hat v_{i,j})\chi[x_i=1;{\bf
  0}]\chi[x_j=1;{\bf 0}]$.  Inserting the $\chi$'s from
eqs.~(\ref{chi-eq},\ref{omega-eq}) gives, for
$i+2m\!\le\!j\!\le\!i+l-1$,
\begin{eqnarray}
&&\frac{a_j^{(i)}}{b_j^{(i)}}\prod_{k=j}^{i+l-1}
\left(1\!-\!\frac{a_k^{(0)}}{b_k^{(0)}}\right)=
\exp(-\hat v_{i,j})\,\frac{a_j^{(0)}}{b_j^{(0)}}\,\prod_{k=j}^{i+l-1}
\left(1\!-\!\frac{a_k^{(i)}}{b_k^{(i)}}\right)\,,\nonumber\\[0.2cm]
&&a_j^{(0)}\equiv{\mathbb E}\,[\,x_j\hspace*{-0.5cm}\prod_{k=j-(l-1)}^{j-1}
        \hspace*{-0.4cm}(1\!-\!x_k)]=
      p_j\!-\!\hspace*{-0.4cm}\sum_{k=j-(l-1)}^{j-2m}\hspace*{-0.2cm}p_{j,k}\,,
   \hspace*{0.6cm}a_j^{(i)}\equiv p_{i,j}\,,\label{correl-eq}\\
&&b_j^{(0)}\equiv {\mathbb E}\,[\hspace*{-0.4cm}
      \prod_{\hspace*{0.3cm}k=j-(l-1)}^{j-1}\hspace*{-0.4cm}(1\!-\!x_k)]
=1\!-\!\hspace*{-0.4cm}\sum_{k=j-(l-1)}^{j-1}\hspace*{-0.2cm}
p_k\!+\!\sum_{k=j-(l-1)}^{j-1-2m}\sum_{\hspace*{0.2cm}r=k+2m}^{j-1}p_{k,r}
\,,\hspace*{0.6cm}b_j^{(i)}\equiv p_i-\sum_{k=i+2m}^{j-1}p_{i,k}\,.\nonumber
\end{eqnarray}
Note that the method for determining the two-point correlator
equations can also be applied for deriving higher order correlator
equations. These equations would be needed for more general
interaction potentials.

Having derived the correlator equations, we can view the moments in
the functional $\Omega$ (\ref{func1-eq}) as functions of the $p_s$ and
minimise $\Omega$ with respect to the $p_s$ to obtain the ``structure
equations''. In fact, these equations can be derived more directly by
using the relation $\chi[{\bf 0}]\exp(-\tilde
u_i)=\chi[x_i=1;{\bf 0}]$, which gives \cite{bound-comm}
\begin{equation}
\exp(-\tilde u_i)\prod_{s=i}^{i+l-1}
         \left(1-\frac{a_s^{(0)}}{b_s^{(0)}}\right)=
\frac{a_i^{(0)}}{b_i^{(0)}}\prod_{s=i+2m}^{i+l-1}
\left(1-\frac{a_s^{(i)}}{b_s^{(i)}}\right)\,.
\label{struc-eq}
\end{equation}
(Equation~(\ref{struc-eq}) together with the correlator equations
(\ref{correl-eq}) determines the occupation probabilities $p_s$.)

An explicit solution of the correlator equations (\ref{correl-eq})
can be obtained for $l\!=\!2m+1$, which means that two rods only interact
if they are in contact (we set $\hat v_i\equiv\hat v_{i,i+2m}$ then).
The solution is
\begin{equation}
\label{pij-eq}
p_{i,i+2m}\!=\!\frac{e^{\hat v_i}(1\!-\!\sum_{k=i}^{i+2m} p_k)
+p_i+p_{i+2m}}{2[e^{\hat v_i}-1]}\left[
\left(1+\frac{4[e^{\hat v_i}-1]p_ip_{i+2m}}
           {e^{\hat v_i}(1\!-\!\sum_{k=i}^{i+2m} p_k)\!+\!
p_i\!+\!p_{i+2m}}
\right)^\frac{1}{2}\!-\!1\right].
\end{equation}
After inserting (\ref{pij-eq}) into (\ref{struc-eq}) one also obtains
the structure equations in explicit form.  For $m=1/2$ and $\hat
v_i=-J_i$ in particular, the system is equivalent to the Ising model
and the structure equations simplify to the ones previously derived
\cite{Percus:1977,Tejero:1987,Samaj:1988} by using very
different approaches.

Unfortunately, for $l\!>\!(2m+1)$ we are not able to give an explicit
solution of the correlator equations (\ref{correl-eq}), since then the
correlator equations represent nonlinear coupled difference equations
rather than a system of simple algebraic equations as in
(\ref{pij-eq}). The solution of the difference equations are uniquely
determined by the boundary conditions induced by the two walls.  Thus
the $p_{i,j}$ will depend on the whole set of occupation probabilities
$\{p_1,\ldots,p_M\}$ in contrast to (\ref{pij-eq}), where
$p_{i,i+2m}=f_i(p_i,\ldots,p_{i+2m})$. The best we can do for
$l\!>\!(2m+1)$ is to reduce the problem to an equation for a
single-site function $h_s\equiv 1-a_s^{(0)}/b_s^{(0)}$, which is the
probability that the site $s$ is not occupied by a rod centre under
the condition that the sites $s-(l-1),\ldots,s-1$ are also not
occupied by rod centres. Using $h_s$, one obtains the following set of
equations equivalent to (\ref{correl-eq},\ref{struc-eq})
\begin{eqnarray}
h_s&=&\frac{1\!-\!\sum_{r=s-(l-1)}^s p_r\!+\!
\sum_{r=s-(l-1)}^{s-2m}\sum_{k=r+2m}^s p_{r,k}}
{1\!-\!\sum_{r=s-(l-1)}^{s-1} p_r\!+\!
\sum_{r=s-(l-1)}^{s-1-2m}\sum_{k=r+2m}^{s-1} p_{r,k}}\label{hi1-eq}\\[0.2cm]
p_{i,j}&=&p_i\,\frac{(1\!-\!h_j)e^{-\hat
  v_{i,j}}\prod_{k=j}^{i+l-1}h_k^{-1}}
{1\!+\!\sum_{n=i+2m}^{i+l-1}(1\!-\!h_n)e^{-\hat
  v_{i,n}}\prod_{k=n}^{i+l-1}h_k^{-1}}\label{hi2-eq}\\[0.2cm]
e^{-\tilde u_i}&=&\frac{(1\!-\!h_i)\prod_{s=i}^{i+l-1}h_s^{-1}}
{1\!+\!\sum_{n=i+2m}^{i+l-1}(1\!-\!h_n)e^{-\hat
  v_{i,n}}\prod_{k=n}^{i+l-1}h_k^{-1}}
\label{hi3-eq}
\end{eqnarray}
Solving (\ref{hi3-eq}) for $h_i$, and inserting this solution into
(\ref{hi2-eq}) gives the $p_{i,j}$ in terms of the $p_s$. Inserting
this solution then into (\ref{hi1-eq}) yields a closed system of
equations for the $p_s$.

At this point we like to note that it is easy to generalise the method
to lattice gases with internal degrees of freedom, where the
occupation numbers can assume $q$ different values $x_s^\alpha$,
$\alpha=1,\ldots,q$ (here and in the following the Greek letters have
to be understood as superscripts). To be specific let us consider the
(generalised) Potts model \cite{Potts:1952}, where
$v_{i,j}^{\alpha\beta}=v_j^{\alpha\beta}\delta_{j,i+1}$.  Following
the procedure described above we find (with $p_s^\alpha\equiv{\mathbb
  E}[x_s^\alpha]$, $p_{s-1,s}^{\alpha\beta}\equiv{\mathbb
  E}[x_{s-1}^\beta x_s^\alpha]$, $p_s\equiv\sum_{\alpha=1}^q
p_s^\alpha$):
\begin{eqnarray}
\Omega&=&\sum_{s=1}^M\Biggl\{
\sum_{\alpha=1}^q \tilde u_s^\alpha p_s^\alpha+
\sum_{\alpha,\beta=1}^q v_s^{\alpha\beta} p_{s-1,s}^{\alpha\beta}
\nonumber\\
&&\phantom{\sum_{s=1}^M\Biggl\{}\hspace*{-0.5cm}+
\sum_{\beta=1}^q \left[ 
   \sum_{\alpha=1}^q p_{s-1,s}^{\alpha\beta}
  \log\frac{p_{s-1,s}^{\alpha\beta}}{p_{s-1}^\beta}+ 
    \Bigl(p_{s-1}^\beta\!-\!\sum_{\alpha=1}^q p_{s-1,s}^{\alpha\beta}\Bigr)
      \log\Bigl(1\!-\!\frac{\sum_{\alpha=1}^q p_{s-1,s}^{\alpha\beta}}
         {p_{s-1}^\beta}\Bigr)\right]\nonumber\\
&&\phantom{\sum_{s=1}^M\Biggl\{}\hspace*{-0.5cm}+
\sum_{\alpha=1}^q\Bigl(p_s^\alpha\!-\!\sum_{\beta=1}^q
p_{s-1,s}^{\alpha\beta}\Bigr)\log\Bigl(\frac{p_s^\alpha\!-\!\sum_{\beta=1}^q
p_{s-1,s}^{\alpha\beta}}{1-p_{s-1}}\Bigr)\nonumber\\
&&\phantom{\sum_{s=1}^M\Biggl\{}\hspace*{-0.5cm}+
\Bigl(1\!-\!p_{s-1}\!-\!p_s\!+\!\sum_{\alpha,\beta=1}^q 
p_{s-1,s}^{\alpha\beta}\Bigr)\log\Bigl(1\!-\!
\frac{p_s\!-\!\sum_{\alpha,\beta=1}^q p_{s-1,s}^{\alpha\beta}}
{1\!-\!p_{s-1}}\Bigr)
\Biggr\}\,,
\label{pottsfunc-eq}
\end{eqnarray}
and for the correlator equations,
\begin{equation}
p_{s-1,s}^{\alpha\beta}=e^{-v_s^{\alpha\beta}}\frac{\left(p_s^\alpha-
\sum_{\gamma=1}^q p_{s-1,s}^{\alpha\gamma}\right)\left(p_{s-1}^\beta
-\sum_{\delta=1}^q p_{s-1,s}^{\delta\beta}\right)}
{\left(1-p_{s-1}-p_s+\sum_{\gamma,\delta=1}^q p_{s-1,s}^{\delta\gamma}
\right)}\,.
\label{pottscorrel-eq}
\end{equation}
The structure equations read
\begin{equation}
e^{-\tilde u_s^{\alpha}}=\frac{(1-p_s)\left(p_s^\alpha-
\sum_{\gamma=1}^q p_{s-1,s}^{\alpha\gamma}
\right)\Bigl(p_s^\alpha-\sum_{\delta=1}^q p_{s,s+1}^{\delta\alpha}\Bigr)}
{p_s^\alpha
\left(1-p_{s-1}-p_s+\sum_{\gamma,\delta=1}^q p_{s-1,s}^{\delta\gamma}
\right)
\left(1-p_s-p_{s+1}+\sum_{\gamma,\delta=1}^q p_{s,s+1}^{\delta\gamma}
\right)}\,.
\label{pottsstruc-eq}
\end{equation}
For $v_s^{\alpha\beta}=v^{\alpha\beta}$ independent of the position
$s$ these equations reduce to the ones previously derived by Percus
\cite{Percus:1982b}.

Finally, we discuss the continuum limit of eq.~(\ref{func1-eq}) for
the pair-potential in eq.~(\ref{v-eq}) and the associated
eqs.~(\ref{hi1-eq}-\ref{hi3-eq}). The continuous system of size $L$ is
composed of hard rods of length $\sigma$ that interact via a potential
$\hat v(x,y)$ ($y\ge x+\sigma$) exhibiting a finite range
$\kappa<2\sigma$.  With $\rho(x)$ being the density of rod centres and
$\rho(x,y)$ the two-point density, we obtain \cite{cont-comm}
\begin{eqnarray}
\Omega&=&\int_0^L\,dy\, \rho(y)\,\tilde u(y)+
\int_0^{L-\sigma}\hspace*{-0.4cm}dy
\int_\sigma^\kappa\,dx\, \hat v(y,y+x)\,\rho(y,y+x)\nonumber\\
&&+\int_0^L dy \Biggl\{
\Bigl[\rho(y)-\int_{y-\kappa}^{y-\sigma} dx \rho(x,y)\Bigr]
\Bigl[\log\Biggl(\frac{\rho(y)-\int_{y-\kappa}^{y-\sigma} dx \rho(x,y)}
{1\!-\!\int_{y-\kappa}^y dx
\rho(x)\!+\!\int_{y-\kappa}^{y-\sigma} dx 
       \int_{x+\sigma}^y dz \rho(x,z)}\Biggr)\!-\!1\Bigr]\Biggr\}\nonumber\\
&&+\int_0^L dy \int_{y-\kappa}^{y-\sigma} \hspace*{-0.5cm} dx\, \rho(x,y)
\Bigl[\log\Biggl(\frac{\rho(x,y)}
             {\rho(x)-\int_{x+\sigma}^y dz
               \rho(x,z)}\Biggr)\!-\!1\Bigr]\,.
\label{func2-eq}
\end{eqnarray}

The set of eqs.~(\ref{hi1-eq}-\ref{hi3-eq}) becomes
\begin{eqnarray}
h(y)&=&\frac{\rho(y)-\int_{y-\kappa}^{y-\sigma} dz\,\rho(z,y)}
{1-\int_{y-\kappa}^y dz\,\rho(z)+\int_{y-\kappa}^{y-\sigma} dx 
\int_{x+\sigma}^y dz\,\rho(x,z)}\,,\label{hy1-eq}\\[0.2cm]
\rho(x,y)&=&\frac{\rho(x)h(y)
         \exp[-\hat v(x,y)]\exp\left(\int_y^{x+\kappa}dz\, h(z)\right)}
{1+\int_{x+\sigma}^{x+\kappa} dz\, h(z)\exp[-\hat v(x,z)]
\exp\left(\int_z^{x+\kappa} du\, h(u)\right)}\,,\label{hy2-eq}\\[0.2cm]
\exp[-\tilde u(y)]&=&\frac{h(y)\exp\left(\int_y^{y+\kappa} dz\,
    h(z)\right)}
{1+\int_{y+\sigma}^{y+\kappa}dz\,h(z)\exp[-\hat v(x,z)]
\exp(\int_z^{y+\kappa} du\, h(u))}\,.\label{hy3-eq}
\end{eqnarray}
Here $h(y)dy$ is the probability that the interval $(y,y+dy)$ is
occupied by a rod centre under the condition that there is no rod
centre in the interval $(y-\kappa,y)$. An explicit solution for $h(y)$
and hence for $\rho(x,y)$ can readily be obtained for the sticky-core
fluid \cite{Baxter:1968}, and inserting this solution into
(\ref{func2-eq}) yields the functional derived in \cite{Percus:1982a}.

The functional in eq.~(\ref{func2-eq}) contains both the density and
the density correlations, and to express the density correlations in
terms of the density $\rho(x)$ one has to solve a nonlinear integral
equation for $h(x)$. This form of the functional reflects the ad-hoc
type of ansatz used in the weighted density functional formalism
\cite{Curtin/Ashcroft:1985}, which turned out to be one of the most
successful methods in the theory of inhomogeneous fluids. The smoothing
of the density on a local scale by a ``weighting function'' is in
particular important to describe more subtle properties, as, for
example, the wetting behaviour in the presence of confining walls
\cite{Dietrich:1988}.

It would be desirable to establish a systematic procedure for deriving
density functionals in higher dimensions $d>1$. The method presented
in this work can straightforwardly be generalised to higher dimensions
as will be shown in forthcoming work.  To keep the problem tractable,
some approximations are needed in $d>1$, but the resulting functionals
will by construction satisfy the requirement of becoming exact upon
dimensional reduction to $d=1$. In comparison to the methods applied
in previous work \cite{Percus:1990}, the procedure based on the Markov
chain has the advantage that the functionals are obtained directly and
no functional integration of the structure equations is needed.

We gratefully acknowledge financial support from the Deutsche
Forschungsgemeinschaft (SFB~513 and Ma~1636/2-1).


\begin{references}
\bibitem{naive-comm} An alternative way to derive ``correlator
  equations'' appears to be a minimisation of $\Omega$ with respect to
  all $p_{k_1,\ldots,k_l}$. Indeed, for the Takahashi model, this
  method reproduces the exact results (see eq.~(\ref{correl-eq})).
\bibitem{bound-comm} For points $i$ (or $j$) near the boundary, i.e.
  $2m\!\le\! i\!<\!l$ (or $M+1-a<j\le M+1-2m$) there would be an
  additional factor $\exp(-\hat v_{0,i})$ (or $\exp(-\hat
  v_{j,M+1})$) due to the fixed rods at the wall. In the correlator
  equations (\ref{correl-eq}) this factor cancels, while in the
  structure equation (\ref{struc-eq}) one has to replace $\tilde u_i$
  by $\tilde u_i+\hat v_{0,i}$ (or $\tilde u_i+\hat v_{i,M+1}$).
\bibitem{cont-comm} When taking the continuum limit, the
  Boltzmann probability becomes a probability density. Taking this
  into account one obtains $\Omega_{\scriptscriptstyle\rm cont.}=
  \Omega_{\scriptscriptstyle\rm discr.}-\log(\Delta
  y)\sum_{s=1}^M p_s +{\rm o}(\Delta y)/\Delta y$, where $\Delta y$
  is the discretisation length.
\end{references}
\end{document}